\documentclass[pra,twocolumn]{revtex4}

\usepackage{graphicx}

\begin{document}

\title{ Tunable Quantum Fluctuation-Controlled Coherent Spin Dynamics}
\author{Jun Liang Song and Fei Zhou}
\affiliation{
Department of Physics and Astronomy,
The University of British Columbia, Vancouver, B. C., Canada V6T1Z1}

\begin{abstract}
Temporal evolution of a macroscopic condensate of ultra-cold atoms is
usually driven by
mean field potentials, either due to scattering between atoms or
due to coupling to external fields; and
coherent quantum dynamics of this type have been observed in various cold
atom
experiments.
In this article, we report results of studies of a class of
quantum spin dynamics which are purely driven by zero point quantum
fluctuations of spin collective coordinates.
Unlike the usual mean-field coherent dynamics,
quantum fluctuation-controlled spin dynamics or {\em QFCSD}
studied here are very sensitive to variation of quantum
fluctuations
and the corresponding driving potentials induced by zero point motions
can be tuned
by four to five orders of magnitude
using optical lattices.
These dynamics have unique dependence on optical lattice potential
depths and quadratic Zeeman fields.
We also find that thermal fluctuations generally can further enhance the induced potentials although
the enhancement in deep optical lattices is much less substantial than in traps or shallow lattices.
{\em QFCSD} can be potentially used to calibrate quantum fluctuations and
investigate
correlated fluctuations
and various universal scaling properties
near quantum critical points.
\end{abstract}
\date{{\small \today}}
\maketitle

\section{introduction}

When particles such as atoms interact with each other at low temperatures,
very often they exhibit remarkably distinct cooperative behaviors as a result of
symmetry breaking.
One of the most fascinating and distinct consequences is
the possibility of observing quantum dynamics at a macroscopic level
\cite{Anderson84,Leggett83}.
In Bose-Einstein condensates of ultra-cold alkali atoms,
macroscopic quantum phenomena related to coherent matter
waves\cite{Mewes97,Bloch99},
AC Josephson effects\cite{Hall98,Anderson98}, and vortex
lattices\cite{Abo-Shaeer01} have all
been observed. Studies of these phenomena in ultra cold matter
should eventually lead to applications
such as cold-atom-based precise measurements.

Spin correlated macroscopic quantum dynamics have also been a focus
of many cold atom experiments carried out recently.
Spin ordering and spin-relaxation collisions were first investigated in
condensates of sodium atoms by Inouye {\em et al.} and Miesner {\em et al.}
from Ketterle's group at MIT\cite{Stenger98,Miesner99}.
Coherent spin dynamics driven by various mean field interactions
or external fields were
later demonstrated in condensates of hyperfine spin-two rubidium
atoms\cite{Schmaljohann04}, and
hyperfine spin-one rubidium atoms\cite{Chang05,Higbie05}.
Ordering in spinor gases is usually induced by hyperfine spin
dependent two-body scattering\cite{Ho98,Ohmi98}.
Coherent spin dynamics observed in
experiments are related to the coherent
quantum dynamics explored in solid state superconductors, and
earlier experiments on ultra cold gases of atoms\cite{Hall98,Anderson98}.
They are explicit manifestations of fascinating macroscopic quantum states
and can be potentially applied towards constructing high precision
interferometers. Remarkably,
coherent dynamics also provide a unique direct measure of interaction
energies or scattering lengths
as emphasized before \cite{Hall98,Schmaljohann04,Chang05}.

Quantum fluctuation-controlled spin dynamics or
{\em QFCSD} we are going to study in this article on the other hand are a direct
measure of quantum fluctuations; they
can be potentially used to calibrate quantum fluctuations and investigate
correlated fluctuations near quantum critical points or universal scaling properties.
Furthermore, {\em QFCSD} of cold atoms can be designed to simulate many other
quantum-fluctuation-induced phenomena such as
Coleman-Weinberg mechanism of spontaneous symmetry breaking\cite{Coleman74}, and
order due to disorder in antiferromagnets\cite{Shender82,Henley89}.
The system we are examining to understand
{\em QFCSD}
is a condensate of rubidium atoms
($^{87}Rb$) in hyperfine spin-two ($F=2$) states.
The two-body
scattering lengths between rubidium atoms have been estimated using
both photoassociation data\cite{Tsai97,Courteille98,Wynar00,Kempen02}, and
elastic scattering data near Feshbach
resonances\cite{Roberts98,Klausen01}.
Most recently, Rabi oscillations
between different two-spin states have
also been used to measure spin dependent interactions\cite{Widera05}.

Quantum fluctuations of spin collective coordinates
at least have three different effects on
spin dynamics. Firstly,
quantum fluctuations of wave lengths of condensate sizes
usually result in finite-size quantum symmetry restoring.
For a spinor condensate of a
few million $^{87}Rb$ (F=2) or $^{23}Na$ (F=1) atoms, restoring of
spin rotational symmetry or quantum diffusion of spin orientation
typically occurs at a time scale of a few tens of seconds.
So they are not relevant to $10-100ms$ spin dynamics studied in
large condensates.
The second well-known effect is to renormalize
semi-classical spin dynamics, such as spin wave
velocities. In high dimensions, dominating contributions are usually
from short wave length fluctuations (in optical lattices the shortest wave
length is set by lattice constants).
These spin fluctuations might also result in quantum phase transitions
between spin ordered and disordered states.
For most dilute cold atom
gases, the renormalization of spin dynamics is
perturbative and
negligible (except around critical points).
And in the limit that interests us, the relative amplitude of
spin fluctuations is small.
The third effect is to
induce dynamics
when conventional
mean field dynamics are completely frozen out because of degeneracies in a
submanifold. This is the limit of {\em QFCSD} that we are going to study.
The dominating
contributions in our case are mainly
from quantum fluctuations of an {\em intermediate} wave length
which is comparable to the de Broglie wave length of an atom traveling
with spin wave velocities.

However it is quite challenging to probe QFCSD in traps without
optical lattices, if not impossible.
For dilute gases in the absence of optical lattices,
the effective driving potential induced by quantum fluctuations
is about $10^{-5}pk$ per particle (see section III for more discussions)
because the relative amplitude of fluctuations is very small.
The corresponding dynamics driven by such a small potential are only visible
at a time scale of a few thousand seconds, too slow to be observed in
current
cold atom experiments. In addition, a tiny external magnetic field
of strength $1mG$ can result in a quadratic Zeeman coupling of order of
$6\times 10^{-3}$pk for rubidium atoms which is three orders of magnitude larger
than the induced potential in dilute gases. In most experiments because of noises in lasers,
the effective quadratic Zeeman coupling
can be controlled only up to a uncertainty that is equivalent to
a magnetic field of an order of $1mG$.
This further complicates future experimental studies of
QFCSD. To resolve these difficulties, we propose to enhance the effect of QFCSD using optical lattices.
To vary the amplitude of quantum fluctuations and optimize the effect of
fluctuations, we
study {\em QFCSD} in optical lattices where the optical potential depth
$V$ is a convenient
tunable parameter\cite{Jaksch98,Greiner02,Folling05,Campbell06,Stoferle04}.

The rest of the article is organized as follows.
In section II, we introduce a lattice Hamiltonian to study dynamics of $^{87}Rb$
atoms in optical lattices and discuss the range of parameters we have used
to investigate this phenomenon. In section
III, we present our main numerical results on quantum-fluctuation induced potentials,
frequencies of coherent dynamics, and how potentials and frequencies depend on optical
lattice potential depth. We also study the dynamical stabilities of coherent oscillations when a quadratic Zeeman
coupling is present.
In section IV, we discuss effects of thermal fluctuations and analyze
the potential-depth dependence of
thermal enhancement of induced
driving potential. In section V, we further investigate effects of spin exchange losses and propose how to
observe quantum-fluctuation-controlled spin dynamics within a relatively short
life time of $F=2$ rubidium atoms.
In section VI, we conclude our studies of QFCSD.

\section{Model}

\subsection{Microscopic Hamiltonian for spin-two rubidium atoms in
optical lattices}

In optical lattices, we use the following Hamiltonian that was introduced
previously\cite{Zhou06,Song07},

\begin{eqnarray}
{\cal H}&=& \sum_{k}
\frac{a_L}{2}\left(\hat{\rho}^2_k-\hat{\rho}_k\right) +\frac{b_L}{2}
\left(\hat{\cal F}^2_k-6\hat{\rho}_k\right) + 5
c_L {\cal D}^\dagger_k {\cal D}^{~}_k \nonumber \\
&-&t_L\sum_{<kl>} (\psi^\dagger_{k,\alpha\beta}
\psi^{~}_{l,\beta\alpha} + h.c.) - \sum_k \mu \hat{\rho}_k +q_B {\cal Q}_{zz}.
 \label{Hamiltonian}
\end{eqnarray}
Here $k$ is the lattice site index and $<kl>$ are the nearest
neighbor sites, $\mu$ is the chemical potential and
$t_L$ is the one-particle hopping amplitude,
$q_B$ is the quadratic Zeeman coupling constant.
$a_L$, $b_L$ and $c_L$ are three interaction constant which have been calculated (see subsection B for more
discussions).
The single band Hamiltonian is valid when all interaction constants above are much smaller
than the energy spacing between centers of two lowest bands.

We have employed the traceless symmetric matrix operator
$\psi^\dagger$ that was
introduced previously for the studies of hyperfine spin two rubidium atoms\cite{Zhou06};
components $\psi^\dagger_{\alpha\beta}$, $\alpha,\beta=x,y,z$
are linear superpositions of
five spin-2 creation operators, $\psi^\dagger_{m_F}$, $m_F=0,\pm1,\pm 2$ as given above.
It is advantageous to use this tensor representation if one is interested in
rotational symmetries of condensate wavefunctions, or construction of
rotationally invariant operators.
We use it to analyze collective spin
modes that correspond to small rotations around various axes.
The tensor operator $\psi^\dagger$ is defined in terms of
the usual creation operators $\psi^\dagger_{m_F}$ for five $F=2$ states,

\begin{eqnarray}
&& \psi^\dagger_{\alpha\beta}=\sum_{m_F=0, \pm 1, \pm 2}
{\cal C}_{\alpha\beta}(m_F) \psi^\dagger_{m_F}; \nonumber \\
&& {\cal C}_{xz}( \pm 1) =
i{\cal C}_{xy}( \mp 2)=
\frac{\mp 1}{\sqrt{2}}, \nonumber \\
&& {\cal C}_{xx} (\pm 2)=-{\cal C}_{yy} (\pm 2)=
\frac{{\cal C}_{yz}(\pm 1)}{i}=\frac{1}{\sqrt{2}},
\nonumber \\
&& {\cal C}_{xx}(0)={\cal C}_{yy}(0)=\frac{{\cal C}_{zz}(0)}{-2}=
\frac{-1}{\sqrt{3}}
\label{coeff}
\end{eqnarray}
where
${\cal C}_{\alpha\beta}( m_F)$ is symmetric with respect to $\alpha\beta$;
all other coefficients are zero.
The number operator $\hat{\rho}$,
the dimer or singlet pair creation operator
${\cal D}^\dagger$, the total spin operator $\hat{F}_\alpha$ are
defined as $\hat{\rho}=1/2 tr \psi^\dagger \psi$, ${\cal
D}^\dagger={1}/{\sqrt{40}}tr \psi^\dagger\psi^\dagger$,
$\hat{F}_\alpha=-i\epsilon_{\alpha\beta\gamma}
\psi^\dagger_{\beta\eta}\psi_{\eta\gamma}$.
And the quadratic Zeeman operator ${\cal Q}_{zz}$ is defined as
${\cal Q}_{zz}=tr \psi^\dagger Q \psi$,
and $Q_{\alpha\beta}=\delta_{\alpha z}\delta_{\beta z}$.

\subsection{Range of interaction parameters for the lattice Hamiltonian}

\begin{figure}
\includegraphics[width=\columnwidth]{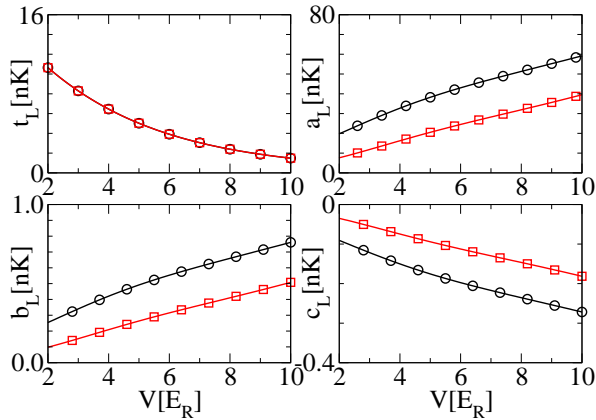}
\caption{(Color online)
Coupling parameters $a_L$, $b_L$, $c_L$ and hoping integral $t_L$
(all in units of $nk$) as a
function of optical potential depth $V$ (in units of recoil
energy $E_R$) in 3D (black circles) and 2D (red squares) optical lattices.
$E_R=3.18 kHz$ for $\lambda=850nm$ lasers.}
\label{parameters}
\end{figure}

Spin correlations between hyperfine spin-two rubidium atoms
are determined by three two-body s-wave scattering lengths $a_F$, $F=0,2,4$.
In optical lattices,
local spin-dependent interactions contain two contributions as
shown in Eq.(\ref{Hamiltonian});
one is, $b_L {\cal F}_k^2/2$,
the energy of having total hyperfine spin ${\cal F}_k$ at site $k$,
and the other is the energy of creating spin singlet pairs ({\em dimers}),
$5c_L {\cal D}_k^\dagger{\cal D}_k$ where ${\cal D}_k$ is the dimer
creation operator\cite{Zhou06}.
The usual contact interaction at site $k$ is of the form
 $a_L (\rho_k^2-\rho_k)/2$, where $\rho_k$ is the number of atoms.
Three effective coupling constants
$a_L, b_L,c_L$ which characterize various interactions are
functions of two-body scattering lengths $a_F$, $F=0,2,4$ and
on-site orbitals $\psi_0(\bf r)$,

\begin{eqnarray}
&& a_L (b_L, c_L) =a(b, c)\frac
{4\pi\hbar^2}{m}
\int d{\bf r} (\psi_0^*({\bf r})\psi_0({\bf r}))^2 .
\label{abc}
\end{eqnarray}
Here $a=(4a_2+3a_4)/7$,
$b=(a_4-a_2)/{7}$ and
$c=(7a_0-10a_2+3a_4)/{35}$ are three effective scattering lengths;
$\psi_0$ is the localized Wannier function obtained by
solving the Schrodinger equation for an atom
in periodical potentials.
$a_L, b_L, c_L$ can then be calculated using the estimates of scattering
lengths obtained in Ref.\cite{Klausen01}.
The range of these parameters for rubidium atoms in optical lattices
is plotted in Fig.\ref{parameters}.

The range of lattice potential depth is chosen to be from zero up to
ten recoil energy or $10 E_R$. ($E_R=h^2 /2m\lambda^2$, where $\lambda$ is the wavelength of
lasers.)
Below $2E_R$, quantum fluctuations turn out to be
too weak to induce substantial potential and dynamics.
For this reason, we only show numerical results for $V$
larger than $2E_R$ but less than $10E_R$.
Within this range, we find that quantum depletion is usually small, less than twenty percent
and our lowest order calculation should suffice.
Although quantum fluctuations can be further enhanced above $10E_R$ and
a superfluid-Mott phase transition should take
place at $13E_R$\cite{Jaksch98,Greiner02}, close to a critical region we
however expect a perturbative calculation like the one
carried out in this article becomes invalid.
The range of atom number density here, or the number of atoms per lattice site $M$
is from zero to three. Most of data are shown for typical values $M=1.0, 2.0, 3.0$.

\begin{figure}
\includegraphics[width=\columnwidth]{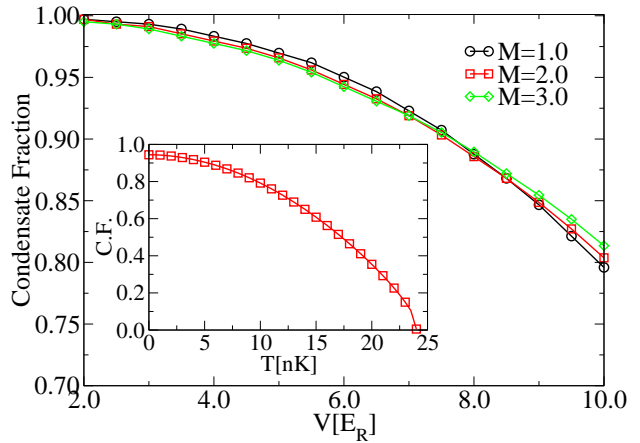}
\caption{(Color online)
Condensation fraction versus optical lattice potential depth $V$ (in units of $E_R$).The inset
shows the condensate fraction at different temperatures when $V$ is
equal to $6E_R$.}
\label{qfccf}
\end{figure}

For $F=2$ rubidium atoms, the quadratic Zeeman coupling $q_B$ is
related to a uniform external magnetic field $B$ via
$q_B=3(\mu_B B/2)^2/\Delta $, $\Delta=6.8GHz$.
Here $\mu_B$ is the Bohr magneton and $\Delta$ is the hyperfine splitting.
For the purpose of studying {\em QFCSD},
we set the range of $B$ to be $70mG > B>1mG$
where effects of quantum fluctuations are most visible; beyond $70mG$ the dynamics are mainly driven by the mean
field quadratic coupling $q_B$ (see section VI for more discussions).
The corresponding range for the quadratic Zeeman coupling is then from
$10^{-2}pk$ to $100pk$.

\subsection{Mean field ground states: quantum spin nematics}

The coherent spin dynamics of a {\em uniform} condensate can be
described by the evolution of a condensate wavefunction
$\tilde{\chi}$, i.e. the expectation value of matrix operator $\psi^\dagger$.
The corresponding equation of $\tilde{\chi}$ is

\begin{eqnarray}
\frac{i}{2}\frac{\partial \tilde{\chi}}{\partial t} &=&
\frac{\partial H_{sc}}{\partial \tilde{\chi}^*}+
\frac{\partial H_{Q}}{\partial \tilde{\chi}^*}+
\frac{\partial H_{qf}}{\partial \tilde{\chi}^*}; \nonumber \\
\frac{\partial H_{sc}}{\partial \tilde{\chi}^*} &=&
\frac{a_L}{4} \tilde{\chi}\textrm{Tr}(\tilde{\chi}^*\tilde{\chi})
+\frac{c_L}{4} \tilde{\chi}^* \textrm{Tr}(\tilde{\chi}\tilde{\chi}) \nonumber \\
 &+& \frac{b_L}{2} [\tilde{\chi}, [\tilde{\chi}^*, \tilde{\chi} ] ] -
(zt_L +\mu/2) \tilde{\chi},
\nonumber\\
\frac{\partial H_{Q}}{\partial \tilde{\chi}^*} &=&q_B  Q \tilde{\chi}.
\label{eom}
\end{eqnarray}
Here $H_{sc}$ is the semiclassical Hamiltonian (of matrix $\tilde{\chi}$, $\tilde{\chi}$)
obtained previously\cite{Zhou06,Semenoff07,Song07},
$H_Q$ is the quadratic Zeeman coupling term with $q_B$ being the coupling
strength. $H_{sc}$ and $H_Q$(per lattice site) are given by:
\begin{eqnarray}
H_{sc} &=& \frac{a_L}{8} \textrm{Tr}(\tilde{\chi}^*\tilde{\chi})
\textrm{Tr}(\tilde{\chi}^*\tilde{\chi})
+\frac{c_L}{8} \textrm{Tr}(\tilde{\chi}^*\tilde{\chi}^*)
\textrm{Tr}(\tilde{\chi}\tilde{\chi})  \nonumber \\
&+&\frac{b_L}{4}\textrm{Tr}[\tilde{\chi}^*, \tilde{\chi}]^2
- (zt_L+\mu/2) \textrm{Tr}(\tilde{\chi}^*\tilde{\chi}); \\
H_Q &=& q_B \textrm{Tr}(\tilde{\chi}^*Q\tilde{\chi}).
\end{eqnarray}
$z$ is the coodination number of optical lattices.
For a field along the $z$-direction, matrix $Q$ is defined as
$Q_{\alpha\beta}=\delta_{\alpha z}\delta_{\beta z}$.
And $H_{qf}$ is the quantum-fluctuation-induced
Hamiltonian discussed below.

For rubidium atoms,
scattering lengths estimated
in Ref.\cite{Courteille98,Wynar00,Kempen02}
lead to $b_L=-10 c_L$ while results
in Ref.\cite{Roberts98,Klausen01} yield $b_L=-2.8 c_L$.
Both calculations show that interaction parameters satisfy
$c_L < 0$ and $4 b_L > c_L$.
As pointed out before\cite{Zhou06}, in this parameter region
without quadratic Zeeman coupling,
ground states are spin nematics characterized by real and symmetric
tensor wavefunctions
$\tilde{\chi}$ (up to an overall phase).
Any condensate that is initially prepared in this submanifold
has no mean field dynamics because the potential gradient $\partial H_{sc}/\partial \tilde{\chi}^*$
vanishes.

To highlight the structure of degenerate ground states,
we consider an arbitrary condensate amplitude $\tilde{\chi}$
in the spin nematic submanifold.
It can be parameterized using
an $SO(3)$ rotation,
${\cal R}$ and a $U(1)$ phase shift
$\phi$, and a real
diagonal traceless matrix $\chi$; i.e.,
\begin{eqnarray}
&& \tilde{\chi}(X_x,X_y,X_z,\xi,\phi)=\sqrt{4M} e^{i\phi} {\cal{R}^T}
\chi(\xi) {\cal{R}}, \nonumber \\
&& {\cal R}(X_x, X_y, X_z)=\exp(T^x X_x+T^y X_y + T^z X_z)
\end{eqnarray}
where $M$ is the number density or average number of atoms per lattice
site.
${\cal R}$ is an $SO(3)$ rotation matrix defined by three spin
angles $X_\alpha$, $\alpha=x,y,z$,
and antisymmetric generators $T^\alpha_{\beta\gamma}=-\epsilon_{\alpha\beta\gamma}$.
$\chi(\xi)$, $\xi\in [0,2\pi]$, are normalized
real diagonal traceless matrices that form a family of solutions
specified by a single parameter $\xi$\cite{Song07};

\begin{eqnarray}
\chi_{\alpha\alpha}=\frac{\sin(\xi-\xi_\alpha)}{\sqrt{3}},
\label{xi-definition}
\end{eqnarray}
and $\xi_x=\pi/6$, $\xi_y=5\pi/6$ and $\xi_z=3\pi/2$.
Nematics with different $\xi$ exhibit different
spin configurations.
Following the definition of tensor operator in Eq.(\ref{coeff}),
one can easily show that these solutions represent
condensates of spin-two atoms specified by five-component
wavefunctions
$\psi^T=(\psi_2,\psi_1,\psi_0,\psi_{-1},\psi_{-2})$ and

\begin{equation}
\psi^T=\sqrt{M}
(\frac{\sin\xi}{\sqrt{2}},0,\cos\xi,0,\frac{\sin\xi}{\sqrt{2}}).
\end{equation}
So in a more conventional representation, these states labeled by value
$\xi$ correspond to
condensates where all atoms occupy a particular spin-2 state,

\begin{eqnarray}
|\xi> = \cos\xi |2,0>+\frac{\sin\xi }{\sqrt{2}} (|2,2>+|2,-2>).
\end{eqnarray}

It is worth pointing out that spin nematics here are time-invariant; the expectation value of the hyperfine
spin operator ${\cal F}_k$
in these states is
zero. However, all nematics have the following nonzero quadrupole spin order (up to an $SO(3)$ rotation),

\begin{eqnarray}
&& {\cal O}_{\alpha\beta}=< {\cal F}_{k\alpha} {\cal F}_{k\beta}> -\frac{1}{3}\delta_{\alpha\beta}
<{\cal F}_k^2>, \nonumber\\
&& {\cal O}_{\alpha\beta}=2 M \sin(2\xi +\xi_\alpha)\delta_{\alpha\beta}.
\end{eqnarray}

Therefore, the nematic submanifold is effectively a five-dimension space that
is characterized by five collective
coordinates $X_\nu$, $\nu=x,y,z,\xi,p$: three spin rotational angles $X_{\alpha}$
($\alpha=x,y,z$), one spin deformation $\xi$-angle $X_{\xi}$ (or $\xi$)
specifying spin configurations and
one phase angle $X_p(=\phi)$.
Furthermore, it is easy to verify that
up to an $SO(3)$ rotation and a phase factor,
states at $\xi$ and at $\xi+\pi/3$ are equivalent and the fundamental
period of this characterization is $\pi/3$.
When the system is rotationally invariant (i.e. no external fields),
we also find that $E(\xi) = E(\xi+\pi/3)$ and
$E(\xi)=E(-\xi)$.
$E(\xi)$ is the energy of a state defined by
$\xi$.

\begin{figure}
\includegraphics[width=\columnwidth]{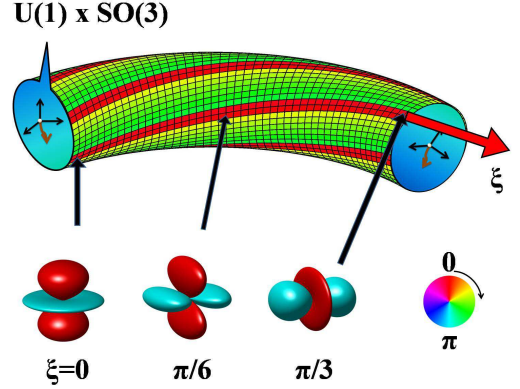}
\caption{(Color online)
An artistic view of the five-dimension manifold of spin nematics.
Three $SO(3)$ rotation degrees and $U(1)$-phase degree are represented by
a cross section at a given spin deformation angle $\xi$.
We have also plotted the nematic wavefunctions
$\psi_S(\theta,\phi)=\sum_{\alpha\beta}\chi_{\alpha\beta}
{n}_\alpha{n}_\beta$ as a function $\theta,\phi$ in a spherical
coordinate.
$n_\alpha$, is the $\alpha$th component of
a unit vector ${\bf n}(\theta,\phi)$
and colors indicate phases of wavefunctions.}
\label{spinwf}
\end{figure}

\subsection{Effective Hamiltonian for QFCSD}

To investigate the kinetic energy and quantum fluctuation-induced
potential energy for dynamics of a condensate
initially prepared in this
submanifold,
we expand tensor $\psi^\dagger_{\alpha\beta}$ about a reference condensate wavefunction
${\chi}(\xi)$
in terms of five collective coordinates $X_\nu$ introduced above
and their conjugate operators $\hat{P}_\nu$.
We furthermore separate the macroscopic dynamics of
condensates (${\bf q}=0$-mode)
from the microscopic zero point quantum fluctuations (${\bf q} \neq 0$ mode).
And we restrict ourselves to
the dynamics of a condensate in a linear regime.

In the {\em appendix},
using the decomposition introduced in Eq.(\ref{Decomposition})
we expand the Hamiltonian in Eq.(\ref{Hamiltonian}) in terms of
collective coordinates $X_\nu$, $\hat{P}_\nu$ and
obtain an effective Hamiltonian for submanifold dynamics.
Up to the quadratic order, the Hamiltonian contains two
sectors, one involving
operators of ${\bf q}=0$-mode and the other one only involving
operators
of ${\bf q}\neq 0$.
The
Hamiltonian for the ${\bf q}=0$ sector
generates the kinetic energy
needed for the dynamics along five orthogonal directions.
The corresponding effective masses
can be expressed in terms of
scattering lengths $a_{0,2,4}$, and quadratic Zeeman coupling $q_B$.
In addition,
the expansion of
the quadratic Zeeman term
$H_{Q}$ for ${\bf q}=0$ mode
also generates a mean field potential $V_{Q}$.
This potential $V_Q$ as illustrated below always favors a biaxial nematic
with $\xi=\pi/2$.
The biaxial nematic has
dihedral-four ($Dih_4$) symmetries with easy
axes in the $xy$-plane\cite{Song07}.
On the other hand, the sector of Hamiltonian for
${\bf q} \neq 0$ collective modes contains zero point energies of those modes.
These energies in general depend on spin configurations or the values of $\xi$.
So quantum fluctuations of collective coordinates effectively
induce a $\xi$-dependent potential $V_{qf}$ in the submanifold;
this potential alone selects out a unique ground state as recently pointed out by Song {\em et
al} \cite{Song07} and Turner {\em
et al} \cite{Turner07}.
For rubidium atoms with a positive $b_L$, previous calculations show that
the ground state is a uniaxial nematic in the absence of quadratic Zeeman coupling.
Turner {\em et al} also pointed out that
thermal fluctuations further enhance the amplitude of
induced potentials and this order-from-disorder phenomenon is robust
against finite temperatures\cite{Turner07}.

We now study the dynamical consequences of both quadratic Zeeman
coupling $H_Q$ and zero point quantum fluctuations $H_{qf}$. For simplicity,
we are mainly focused on dynamics around

a) {\em a uniaxial nematic at $\xi=0$} or a condensate with rubidium atoms
occupying hyperfine spin state $|2,0>$;

b) {\em a biaxial nematic at $\xi=\pi/2$} or a condensate with rubidium
atoms occupying hyperfine spin state
$(|2,2>+|2,-2>)/\sqrt{2}$.

The resultant Hamiltonian for oscillations around
a state $\xi$ ($\xi=0$ or $\pi/2$) can
be cast in the following form,

\begin{eqnarray}
{H_{qf}}&=&\sum_\nu \frac{\hat{P}_\nu^2}{2 N_T m_\nu} + N_T \big(V_{qf}(\xi, X_{\xi}) +V_Q(\xi, X_{\nu}) \big), \nonumber \\
V_{qf}(\xi, X_\xi)&=&\frac{1}{2 N_T}\sum_{\mathbf{q}\neq 0}\sum_{\nu} E_{\nu,\bf q}(\xi,X_{\xi},q_B),\nonumber \\
V_Q(\xi, X_{\nu})&=& \frac{4Mq_B}{3} \cos^2 (\xi+X_\xi) +
4Mq_B\sum_{\alpha}  \kappa_{\alpha}  X^2_\alpha.
\label{effectiveH}
\end{eqnarray}
In Eq.(\ref{effectiveH}), $N_T$ is the number of lattice sites
and $M$ is the average number of atoms per site. The masses
for five directions are calculated to be
\begin{eqnarray}
m_p&=&\frac{1}{a_L+c_L},m_{\xi}=\frac{2M}{ q_B \kappa_\xi -2M
c_L},
\nonumber \\
m_\alpha &=&\frac{8 M G_{\alpha\alpha}}{8M b_L G_{\alpha\alpha}-2M c_L
+\kappa_\alpha q_B G^{-1}_{\alpha\alpha}}.
\label{mass}
\end{eqnarray}
$\kappa_{x,y,z,\xi}$ are $\xi$-dependent;
$\kappa_{x(y)}=\chi^2_{xx(yy)}-\chi^2_{zz}$,
$\kappa_z=0$ and is
independent of
$\xi$;
$\kappa_{\xi}=4(\dot{\chi}^2_{zz}-\chi^2_{zz})$ is a function of $\xi$
and is $-4/3$ when
$\xi=0$, and $4/3$ when $\xi=\pi/2$; $\dot{\chi}_{zz}=d\chi_{zz}/d\xi$.
$G_{\alpha\alpha}(\xi)$, $\alpha=x,y,z$ is a function of $\xi$,
$G_{xx,yy}(\xi)= \sin^2 (\xi\mp
\frac{2\pi}{3})$, $G_{zz}= \sin^2 \xi$.
$E_{\nu,\bf q}(\xi,X_\xi, q_B)$ is the energy of a mode-$\nu$
($\nu=x,y,z,\xi,p$)
collective excitation
with crystal momentum ${\bf q}$, and is a function of parameters $\xi, X_\xi$, and quadratic
Zeeman coupling $q_B$.
In the {\em appendix}, we show the general form of this energy explicitly for
the case of nonzero $q_B$.
Only spin modes with $\nu=x,y,z,\xi$ contribute to
the $\xi$-dependence of potential $V_{qf}$.
Fluctuations of phase modes are independent of parameter $\xi$ or Zeeman coupling $q_B$
and are irrelevant for discussions of spin dynamics as a result of spin-phase separation.
Note that $V_{qf}$ is a function of $\xi$, $q_B$ and $X_{\xi}$;
and $V_Q$ is a function of $\xi$, $q_B$ and $X_{\nu}$, $\nu=x,y,z,\xi$.

\section{Main Results for QFCSD}

\subsection{Potentials induced by quantum fluctuations}

\begin{figure}
\includegraphics[width=\columnwidth]{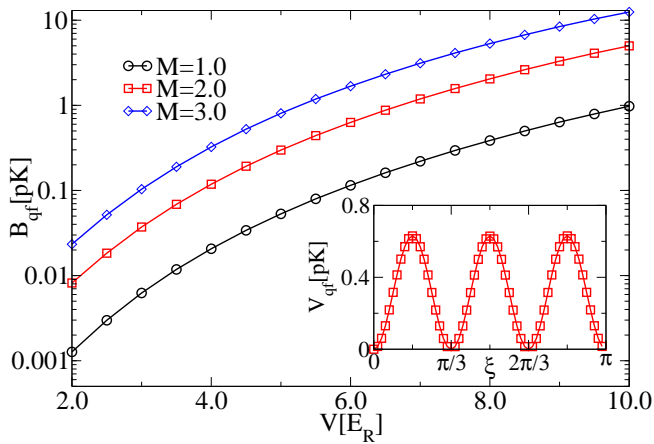}
\caption{(Color online)
Barrier height $B_{qf}$ (in units of pK) as a
function of optical potential depth $V$ (in units of recoil
energy $E_R$). Inset is for the $\xi$-dependence of $V_{qf}$, a potential induced by
quantum fluctuations when $V=6E_R$ and $M=2.0$.}
\label{fig-Barrier}
\end{figure}

We first consider a situation where the quadratic Zeeman coupling is absent
and the potential $V_Q$ vanishes.
As argued before, generally speaking
$V_{qf}(\xi, 0) = V_{qf}(\xi+\pi/3, 0)$ and $V_{qf}(\xi, 0)=V_{qf}(-\xi, 0)$.
The explicit form of $V_{qf}$ calculated here indeed is consistent with
this general requirement. We also find that the main contribution to
$V_{qf}$ is from fluctuations of wavelength $1/(m_{BN} v_\alpha)$ that
is much longer than the lattice distance $d_L$.
More specifically, following discussions in the {\em appendix},
the $\xi$-dependent induced potential can be written as a sum
of polynomials of spin-wave velocities, i.e. $\sum_\alpha
v^{d+2}_\alpha$ for $d$-dimension optical lattices.
Spin-wave velocities are functions of $c_L, b_L$ and $t_L$.
We numerically integrate over all wavelengths and
obtain the $\xi$-dependence of $V_{qf}$ as shown in the inset of Fig.\ref{fig-Barrier}.

Consequently, we find that in d-dimension lattices
($d=2,3$), the barrier height $B_{qf}$ which is defined as
$V_{qf}(\frac{\pi}{6},0)-V_{qf}(0,0)$, the energy difference
between $\xi=\pi/6$ and $\xi=0$
satisfies the following simple scaling function,

\begin{eqnarray}
B_{qf}= \frac{|M
c_L|^{\frac{d+2}{2}}}{t_L^{\frac{d}{2}}}
g_d(\frac{b_L}{c_L}, \frac{M|c_L|}{t_L}).
\label{barrier}
\end{eqnarray}
Here $g_d(x,y)$
is a dimensionless function that can be studied numerically.
This scaling function is either insensitive to the variation of $y$
as for $d=3$ \cite{Song07} or
independent of $y$ as for $d=2$.

We also study the barrier height $B_{qf}$ as a function of $V$,
the potential depth of optical lattices.
We find that in the absence of lattice potentials or in traps, for a density that is equivalent to
one particle ($M=1.0$) per lattice site the barrier height is of
order of $10^{-5}$ to $10^{-4}pk$ and is negligible in experiments.
When the optical lattice potential depth $V$ is varied,
the barrier height typically increases by four or five orders of
magnitude. Particularly,
as $V$ increases from $2E_R$ to $10 E_R$, mean field
interaction energies $a_L, b_L, c_L$ vary by less than a factor of three;
however, for the same range of $V$,
the barrier height $B_{qf}$
varies from $10^{-3}pk$ to a few $pk$.
The energy shift between the uniaxial state at $\xi=0$
and biaxial state at $\xi=\pi/6$
is analogous to the Lamb shift observed in atoms\cite{Lamb47}.
In Fig.\ref{fig-Barrier}, we show the barrier height $B_{qf}$ of induced potential verus
lattice potential depth $V$.

\begin{figure}
\includegraphics[width=\columnwidth]{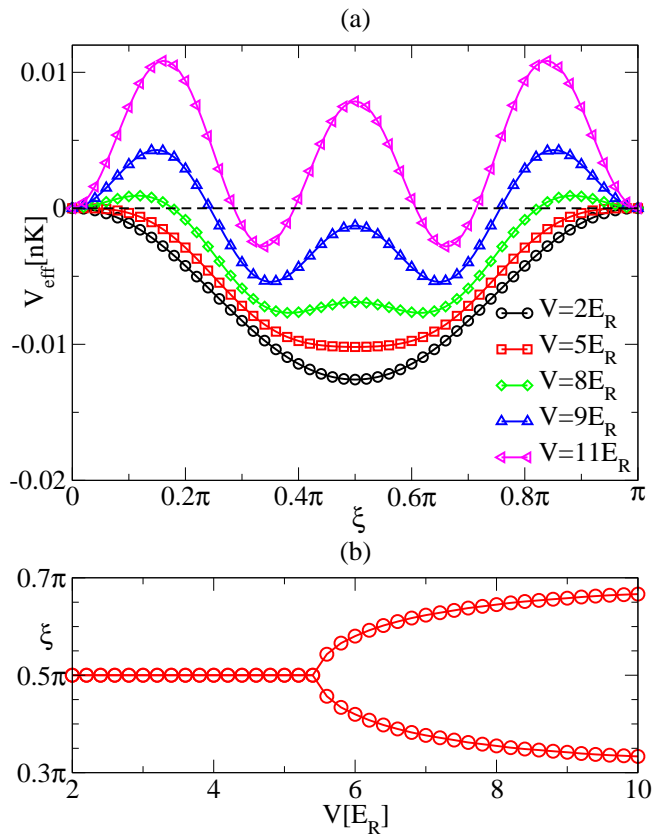}
\caption{(Color online)
(a) Effective potential $V_{eff}=V_Q+V_{qf}$ (in units of $nk$) as a
function of $\xi$ for various optical potential depth $V$ (in units of recoil
energy $E_R$) of 3D lattices. The quadratic Zeeman coupling
strength is set to be $10pk(31mG)$ and number of atoms per lattice site $M$ is equal to one.
(b) Values of $\xi$ at which global potential minima in (a) are found
are plotted as a function of
optical potential depth $V$.}
\label{fig-potent}
\end{figure}

In Fig.~\ref{fig-potent}, we further plot the effective potential
$V_{eff}=V_Q+ V_{qf}$
as a function of optical potential depth $V$. When $V$ is less than $V_c$,
the biaxial nematic with $\xi=\pi/2$ is stable and
as $V > V_c$, it becomes locally unstable.
For $^{87}Rb$ atoms with one atom per lattice site ($M=1.0$),
$V_c$ is about $5.5 E_R$ ($E_R$ is the recoil energy of optical lattices)
when the quadratic Zeeman coupling is $10pk$.
Almost opposite behaviors are found for
uniaxial nematics at $\xi=0$.
Values of $\xi$ at which global potential minima
are found are shown as a
function of $V_{eff}$ in
Fig.~\ref{fig-potent}b.

\begin{figure}
\includegraphics[width=\columnwidth]{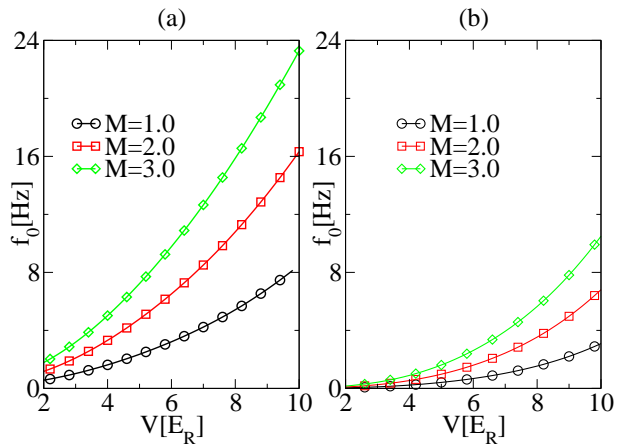}
\caption{(Color online)
(a),(b), frequencies $f_0=\Omega_0/(2\pi)$ for population oscillations around $|2,0>$ state
are shown as a function of potential depth $V$ in (a) 2D and (b)
3D optical lattices for different atom number density $M$;
here the quadratic Zeeman coupling is absent. }
\label{fig-frequency0}
\end{figure}

\subsection{Oscillations induced by quantum fluctuations}

To understand the dynamical consequences of $V_{qf}$,
we consider coherent dynamics
around the uniaxial nematic
state at $\xi=0$ that is selected out by the fluctuation-induced potential
$V_{qf}$ or the biaxial nematic state at $\xi=\pi/2$ that is favored by
the
quadratic Zeeman
potential $V_Q$. Particularly we are interested in oscillations along
the $X_\xi$-direction around state $\xi=0$ or $\pi/2$.
These motions correspond to the following time evolution of condensates,

\begin{eqnarray}
&& |t>\approx ({\cos\xi}- \sin\xi X_0 F(t))
|2,0> + \nonumber \\
&& \frac{1}{\sqrt{2}}(\sin \xi + \cos\xi X_0 F(t)) (|2,2>+|2,-2>);
\nonumber \\
&& F(t) =\cos\frac{\Omega}{2}t +i \frac{m_\xi \Omega}{4M} \sin\frac{\Omega}{2} t
\nonumber \\
\label{population}
\end{eqnarray}
where $X_0$ is the initial deviation from state $\xi=0$ or $\pi/2$
and $m_\xi$ is the effective mass given in Eq.(\ref{mass}).
By solving the equation of motion in Eq.(\ref{effectiveH}) including the
quadratic Zeeman
effects, we derive a general expression for oscillation frequencies
$\Omega (=2\pi f)$.
The oscillation frequency $\Omega(q_B, V)$
in general is a function of
the quadratic Zeeman coupling $q_B$,
optical lattice potential depth $V$, and
the number density $M$.
For population oscillations around $\xi=0$ or $\pi/2$,
we obtain the following frequencies

\begin{eqnarray}
&& \Omega =2 \sqrt{\big(\sigma \frac{8Mq_B}{3} +
V^{''}_{qf} \big)
\big( \sigma
\frac{2q_B}{3M}
+|c_L| \big)};
\label{freq}
\end{eqnarray}
here $\sigma=\mp 1$ for $\xi=0$ and $\pi/2$ respectively.
$V^{''}_{qf}$ is the curvature of potentials at $\xi=0$ or $\pi/2$.

Note that both the real and imaginary part of $F(t)$ oscillate as a function of time;
thus unless $\frac{m_\xi \Omega}{4M}$ is exactly equal to one which
only occurs
when atoms are noninteracting or the quadratic Zeeman coupling $q_B$ is
infinite, the magnitude of $F(t)$
oscillates
as the time varies.
Practically, in most cases we examine below,
$m_\xi \Omega$ turns out to be much less than unity
because both the curvature $V^{''}_{qf}$ and the Zeeman coupling $q_B$
are smaller than the inverse of
effective mass $m_{\xi}$,
or the spin interaction energy $c_L$;
thus oscillations of the imaginary part of $F(t)$ can be neglected.
Oscillations
along the direction of $X_\xi$ therefore always
lead to temporal oscillations in
population of rubidium atoms
in either $|2,0>$ or $|2,\pm 2>$ states that can be observed in experiments.

We first examine the cases when the quadratic Zeeman coupling is absent
i.e. $q_B=0$. The frequency in this limit is a direct measure of
quantum fluctuations and $\Omega_0=\Omega(q_B=0, V)=2\sqrt{V_{qf}^{''}
|c_L|}$.
In the absence of lattice potentials, or in traps
we find that oscillation frequencies are about $10^{-3}$Hz.
At finite temperatures,
thermal fluctuations do enhance oscillation frequencies by a
factor of two to four;
however, this is far from sufficient for the experimental study of {\em
QFCSD} within the
life time of these isotopes\cite{Schmaljohann04}.

In optical lattices, we find that
the enhancement of spin-dependent interactions $b_L, c_L$, and
especially the rapid increasing of band mass $m_{BN}$ can result in
oscillation frequencies of order of a few Hz, which are about three to
four orders of
magnitude higher than those in traps.
The scaling behavior of $\Omega_0$ is closely related to that of $B_{qf}$.
Taking into account the expression for effective mass $m_\xi$, we find

\begin{eqnarray}
\Omega_0 \sim \frac{M^
{\frac{d+2}{4}}
|c_L|^{\frac{d+4}{4}}}{t_L^{\frac{d}{4}}}
g^{\frac{1}{2}}_d(\frac{b_L}{c_L}, \frac{M|c_L|}{t_L}).
\end{eqnarray}
In Fig.~\ref{fig-frequency0}, we show the plot of $f_0=\Omega_0/2\pi$, the
oscillation frequency versus $V$ in the absence of Zeeman coupling $q_B$.

\begin{figure}
\includegraphics[width=\columnwidth]{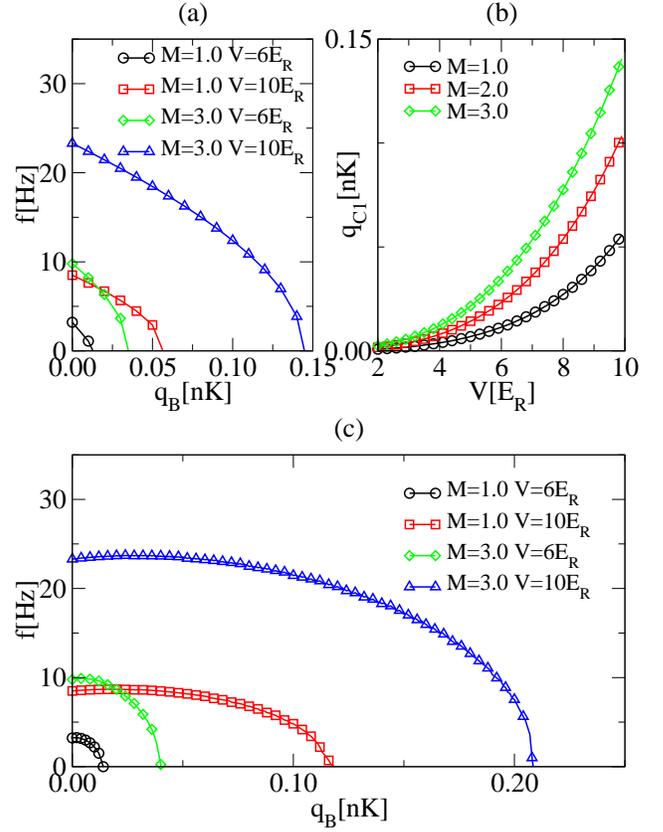}
\caption{ (Color online)
Frequencies $f=\Omega/(2\pi)$ for population oscillations as a function of quadratic
Zeeman coupling $q_B$(in units of $nk$) for various potential depth $V$
in 2D optical lattices with $M$ atoms per lattice site.
(a) for oscillations along the $X_\xi$-direction
around state $|2,0>$; dynamics are unstable above a threshold $q_{c1}(V)$.
The V-dependence of $q_{c1}$ (in units of $nk$) is shown in (b).
(c) is for oscillations around states with $\xi=\xi_m (< \pi/2)$ at which
global potential minima shown in Fig.\ref{fig-potent}(a) are found.
(see also Eq.(\ref{population}))
 }
\label{fig-frequency1}
\end{figure}

\begin{figure}
\includegraphics[width=\columnwidth]{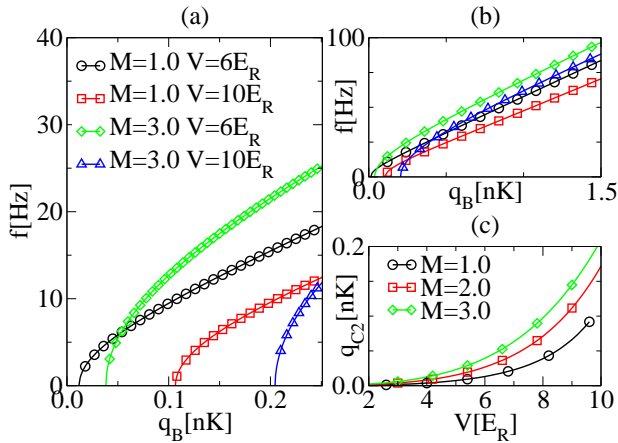}
\caption{ (Color online)
(a), (b) Frequencies $f=\Omega/(2\pi)$ for oscillations along the $X_\xi$-
direction
around biaxial nematic state $|2,2>+|2,-2>$ versus quadratic Zeeman
coupling $q_B$ (in units of $nk$) for various optical potential depth $V$
in 2D optical lattices with $M$ atoms per
lattice site.
(a) is for population oscillations close to the threshold $q_{c2}$
and (b) is for away from $q_{c2}$.
Oscillations are unstable below a threshold $q_{c2}$;
in (c), $q_{c2}$ (in units of $nk$) as a function of $V$ is shown.}
\label{fig-frequency2}
\end{figure}

Experimental studies of {\em QFCSD} can also be carried out by
investigating
dynamics of rubidium atoms in the presence of finite quadratic Zeeman
coupling $q_B$.
The submanifold dynamics
now are driven by both quadratic Zeeman
coupling and quantum fluctuations.
There are three main modifications to the effective Hamiltonian $H_{eff}$.
Firstly, according to Eq.(\ref{mass}),
effective masses now depend on the quadratic Zeeman coupling $q_B$.
Secondly, following discussions in the {\em appendix}, among four spin
collective modes
($\nu=x,y,z,\xi$) and one phase mode ($\nu=p$),
two of spin modes, $x$- and $y$-mode,
can be gapped and one,
$z$-mode, always remains gapless.
Details of spectra of $\xi$-mode depend on the value of $\xi$;
excitations are unstable when $\kappa_{\xi}$ is negative and
are gapped when $\kappa_{\xi}$ are positive.
The energy gap of stable $\xi$-mode excitations depends on $\xi$
and becomes maximal
at $\xi=\pi/2$.
Finally, a mean field quadratic Zeeman potential that
depends on $X_\xi$ and $X_\alpha$, $\alpha=x,y,z$,
is also present and modifies the spin dynamics.
Oscillation frequencies $\Omega$ as a function of
quadratic Zeeman coupling $q_B$ for various potential depth $V$
can be calculated explicitly using Eq.(\ref{freq}).
Plots of these results
are shown in Fig.\ref{fig-frequency1},\ref{fig-frequency2}.

In the limit of weak quadratic Zeeman coupling,
{\em QFCSD}
can be most conveniently studied around
the uniaxial nematic state at $\xi=0$, or $|2,0>$ state\cite{stability}. For rubidium
atoms,
this is the ground state when the quadratic Zeeman coupling is absent,
and remains to be
locally stable along the direction of $X_\xi$ up to a finite $q_{c1}$.
When the quadratic Zeeman coupling $q_B$ is much smaller than $q_{c1}$,
the effective potential $V_{eff}$ is
mainly due to the quantum-fluctuation induced one, $V_{qf}$.
Above $q_{c1}$, a
dynamical instability occurs and
a perturbation along the $X_\xi$-direction around $|2,0>$ starts to grow
exponentially.
We find that in traps where $V=0$, $q_{c1}$ is about
$10^{-2}pk$;
in optical lattices when $V$ varies between $2 E_R$ to $10 E_R$, the
value of $q_{c1}$ increases from $0.1 pk$ up to about $0.1 nk$.
And as $q_B$ approaches zero, the oscillation frequency
$\Omega(q_B, V)$ saturates at the
value $\Omega_0$ shown in Fig.\ref{fig-frequency0}.
The $q_B$-dependence of frequencies for oscillations around $\xi=0$ is
shown in
Fig.\ref{fig-frequency1}; near
$q_B=q_{c1}$, the frequencies scale as $\sqrt{q_{c1}-q_B}$.
Alternatively, one can also study oscillations around
states at which global potential minima of $V_{eff}$ are found;
similar $q_B$-dependence is shown in
Fig.\ref{fig-frequency1}c.

\begin{figure}
\includegraphics[width=\columnwidth]{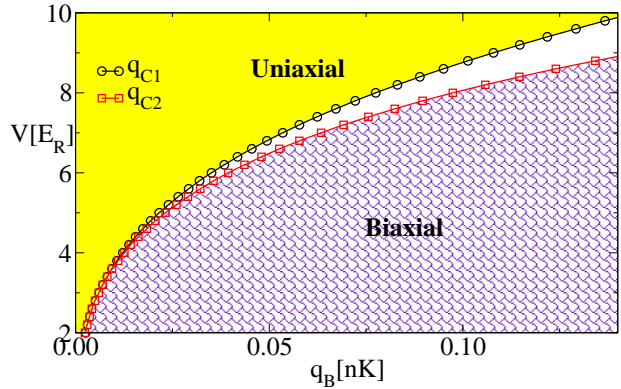}
\caption{ (Color online)
Regions of dynamical stability in the $V-q_B$ plane for three particles per
lattice site ($M=3.0$).
Oscillations around uniaxial nematic or state $|2,0>$ (biaxial nematic or
state $|2,2>+|2,-2>$)
are locally stable along the $X_\xi$-direction in the yellow shaded
(purple patterned) area;
dynamical instabilities occur when $q_B > q_{c1}(V)$ ($q_B< q_{c2}(V)$).
$q_B$ is in units of $nk$.}
\label{fig-frequency3}
\end{figure}

At relatively high frequencies,
convenient oscillations to investigate are the ones along the $X_\xi$
direction
around the biaxial nematic state at $\xi=\pi/2$ point or
$(|2,2>+|2,-2>)/\sqrt{2}$ state.
This state becomes stable when $q_B$ is larger
than $q_{c2}$
and oscillations are well defined only in this limit.
When $q_B$ is smaller than this critical value, the dynamics
are mainly driven by quantum fluctuations and oscillations
are unstable. In the vicinity of $q_{c2}$,
the oscillation frequency again scales as $\sqrt{q_B-q_{c2}}$.
When $q_B$ is much bigger than $q_{c2}$ but smaller than $c_L$,
the frequency is proportional to $\sqrt{q_B}$; in this limit,
the potential $V_{eff}$ is already dominated by the quadratic Zeeman term,
or $V_{Q}$, but the effective mass is mainly induced by spin
dependent interactions.

When the quadratic Zeeman coupling further increases well above the
value of $c_L$, relatively fast dynamics are now mainly driven by the
external
coupling and the frequency $\Omega$ approaches $8q_B/3$, which is equal
to the quadratic Zeeman splitting between state $|2,0>$ and $|2,\pm 2>$.
In this limit, scattering between atoms
during a short period of $2\pi/\Omega$
becomes negligible;
the population oscillations due to scattering or interactions are
therefore significantly suppressed
as here hyperfine spin two atoms are effectively {\em noninteracting}.
Indeed, following the expression for $F(t)$ in Eq.(\ref{population}) and Eq.(\ref{mass}),(\ref{freq}), one
finds that the amplitude of population oscillations scales
as $1/q_B$ at large $q_B$ limit and
$F(t)$ becomes $\exp(i 4 q_B t/3)$ when $q_B$ approaches infinite, i.e., a
pure phase factor with a constant modulus.

In Fig.\ref{fig-frequency2},
we show that the $q_B$-dependence of oscillation frequencies both close to the
threshold $q_{c2}$ and away from $q_{c2}$.
As we have mentioned before, {\em QFCSD} is rather sensitive to the
variation
in $V$. $q_{c1}$ and $q_{c2}$ as functions of $V$ are shown in
Fig.~\ref{fig-frequency1},\ref{fig-frequency2}.

So far we have studied coherent dynamics driven
by potentials induced by quantum fluctuations at zero temperature.
In the following section, we analyze the effect of temperatures, or
thermal fluctuations.

\section{Effects of thermal fluctuations}

We now turn to the effect of temperatures and focus on three-dimensional
optical lattices.
At a finite temperature, collective modes including spin waves are
thermally excited. The occupation number of $\alpha$th spin-wave excitations with momentum
$\hbar q$ and energy $E_{\alpha,q}$ is

\begin{eqnarray}
n(\alpha,q)=\frac{1}{e^{\frac{E_{\alpha, q}}{kT}}-1}.
\end{eqnarray}
The free energy density (or per lattice site) of a condensate characterized by $\xi$ is

\begin{eqnarray}
F(\xi, T, V)=\frac{1}{N_T}\sum_{\alpha, q} \left( \frac{E_{\alpha, q}}{2}
+k T \ln (1-e^{-\frac{E_{\alpha,q}}{kT}})\right).
\end{eqnarray}

Just as in the zero temperature case
because of the $\xi$-dependence of spin wave velocities,
the free energy density also depends on the values of $\xi$
and the main contribution to the $\xi$-dependence of free energy density
is again from fluctuations of wavelength $1/(m_{BN}v_\alpha)$, or
of a characteristic energy $T_*$ that is of an order of
$|c_L|$ (or $b_L$). Thermal fluctuations become more important than
quantum ones
when temperatures are much higher than $T_*$.
The other temperature effect is from the temperature dependence of
spin wave velocities $v_\alpha$. Becuase of the thermal depletion,
spin wave velocities $v_\alpha$($\alpha=x,y,z$) decrease as temperatures
increase and become
zero when the BEC temperature $T_{BEC}$ is
approached.
Therefore the temperature dependence of free energy are determined by
two dimensionless quantities, $T/T_*$ and $T/T_{BEC}$.
In all cases we are going to study below, $T_*$ (of order of $-c_L$) is less or much less than
$T_{BEC}$.

We study the asymptotics
of free energy in low temperature ($T \ll T_*$) and high temperature
($ T \gg T_*$) limits.
When $T \ll T_*$, only modes with energy much smaller than $T_*$
are thermally occupied. Therefore contributions from these thermal
excitations can be calculated by substituting
quasi-particle spectra $E_{\alpha,q}$ approximately
with phonon-like spectra $v_{\alpha} q$. One then obtains the following
expression
for the free energy per lattice site (up to a constant)

\begin{eqnarray}
F(\xi,T, V)&=&V_{qf} (1 - (\frac{T}{T_{BEC}})^{\frac{3}{2}})^{\frac{5}{2}} \nonumber \\
 &-& \frac{\pi^2 kT}{90}
\sum_{\alpha} \left(\frac{d_L kT}{v_\alpha}\right)^3 .
\end{eqnarray}

The barrier height, $B_{th}(T, V)=F(\xi=\frac{\pi}{6}, T, V)-F(\xi=0, T,
V)$,
can be calculated accordingly and the asymptotics at low temperatures is

\begin{eqnarray}
B_{th}(T, V)-B_{th}(0, V)\sim \frac{|c_L|^{5/2}}{t_L^{3/2}}
(\frac{kT}{|c_L|})^4 .
\label{bb}
\end{eqnarray}
Here we have neglected the term depending on $T/T_{BEC}$ because we are interested
in temperatures much less than $T_{BEC}$.(Such a term is only important at very low temperatures
that are of an order of $|c_L|  (|c_L|/T_{BEC})^{3/5}$ $(\ll T_*)$).
Therefore, the low temperature enhancement of $B_{th}$ is mainly
from the "black-body" radiation of spin wave excitations.

When $T \gg T_*$, spin modes with energy much bigger than $T_*$
are thermally occupied. For modes of energy $T_*$, we can approximate
$n(\alpha, q)$ with a classical result $n(\alpha, q)= kT/E_{\alpha, q}$.
Our scaling analysis of the free energy per lattice site shows that in
this limit,

\begin{eqnarray}
F(\xi,T, V)=-\frac {2kT}{3\pi}
\sum_{\alpha} (\frac{v_\alpha}{4 d_L t_L})^3
(1-(\frac{T}{T_{BEC}})^{\frac{3}{2}})^{\frac{3}{2}}.
\label{eq20}
\end{eqnarray}

The barrier height, $B_{th}(T, V)=F(\xi=\frac{\pi}{6}, T, V)-F(\xi=0, T,
V)$,
can be calculated accordingly and asymptotics at high temperatures are

\begin{eqnarray}
B_{th}(T, V)\sim\frac{c_L^{5/2}}{t_L^{3/2}} \frac{kT}{c_L}
(1-(\frac{T}{T_{BEC}})^{\frac{3}{2}})^{\frac{3}{2}}.
\label{barrierth}
\end{eqnarray}
The above result shows that at relatively high temperatures thermal spin wave
fluctuations enhance induced potentials and this enhancement is characterized by
a linear function of $kT/|c_L|$ or $kT/T_*$. However,
because of the thermal depletion of condensates at finite temperatures, there is an overall
suppression given by a factor $1-(T/T_{BEC})^{3/2}$.
The competition between these two effects results in a maxima in the
free energy density.
That is at temperatures larger than $T_*$ but much smaller than $T_{BEC}$,
the free energy increases linearly as a function of temperature
until the thermal depletion
becomes significant. The maximal enhancement is therefore about
$T_{BEC}/T_*$. At further higher temperatures, the free energy density
decreases and
near $T_{BEC}$,
its temperature dependence is mainly determined by a factor
$(1-({T}/{T_{BEC}})^{3/2})^{3/2}$ as shown in Eq.(\ref{barrierth}).

\begin{figure}
\includegraphics[width=\columnwidth]{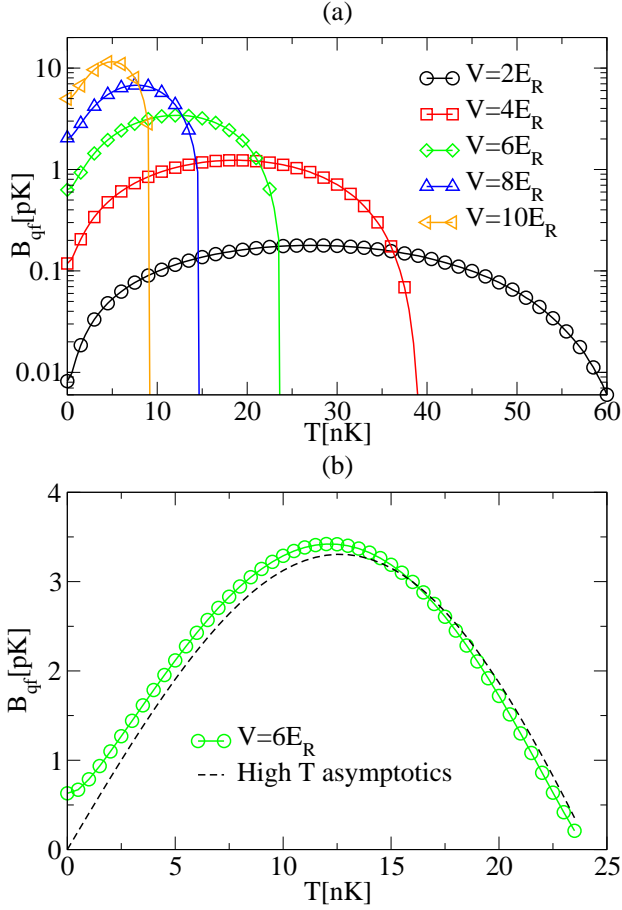}
\caption{(Color online)
Induced free energy barrier height as a function of temperature for different
lattice potentials. The dashed line in (b) is a plot for
the high temperature asymptotics derived
in Eq.(\ref{eq20}). Note that thermal enhancement of barrier height is much less significant in deep
optical lattices.}
\label{fig-BarrierT}
\end{figure}

Evidently,
the magnitude of maximal enhancement is very much dependent
on the ratio between $T_{BEC}$ and $T_*$. In optical lattices,
the magnitude of BEC transition temperatures for averagely one or two
particles per lattice is mainly
set by the band width $t_L$ while $T_*$ is determined by $|c_L|$.
This ratio therefore the enhancement is quite sensitive to
the optical lattice potential depth $V$.
We find that the enhancement in the absence of lattice potentials is substantial and thermal fluctuations
increase the induced potential by about a factor of fifty. This is consistent with
a previous calculation\cite{Turner07}. However, when the lattice potential
depth increases,
$c_L$ increases but $t_L$ decreases, the thermal enhancement becomes much less
significant. At $V=10E_R$, we find that the thermal effect only increases the potential
by a few tens of percent and the potential can be predominately due to quantum fluctuations.

\section{Effects of spin exchange losses}

For $F=2$ rubidium atoms, the life time is mainly due to spin exchange losses that take place at a relatively
short time scale
of about 200ms\cite{Schmaljohann04, Chang05}. Note that three-body recombination usually occurs at a much longer time scale and has little effect on
the time evolution studied here.
Spin exchange losses put a very serve constraint on possible observations
of {\em full} coherent oscillations driven by quantum
fluctuations at frequencies well below 5Hz.
To overcome this difficulty, we suggest to
apply a quadratic Zeeman coupling close to $q_{c1}$ ($q_{c2}$)
and study time evolution of state $|2,0>$ or
$\frac{1}{\sqrt{2}}(|2,2>+|2,-2>)$.
Regions of dynamical stability of these states
can be obtained by studying small oscillations around them as shown in section III.
Results are summarized in
Fig.\ref{fig-frequency3}.

\begin{figure}
\includegraphics[width=0.98\columnwidth]{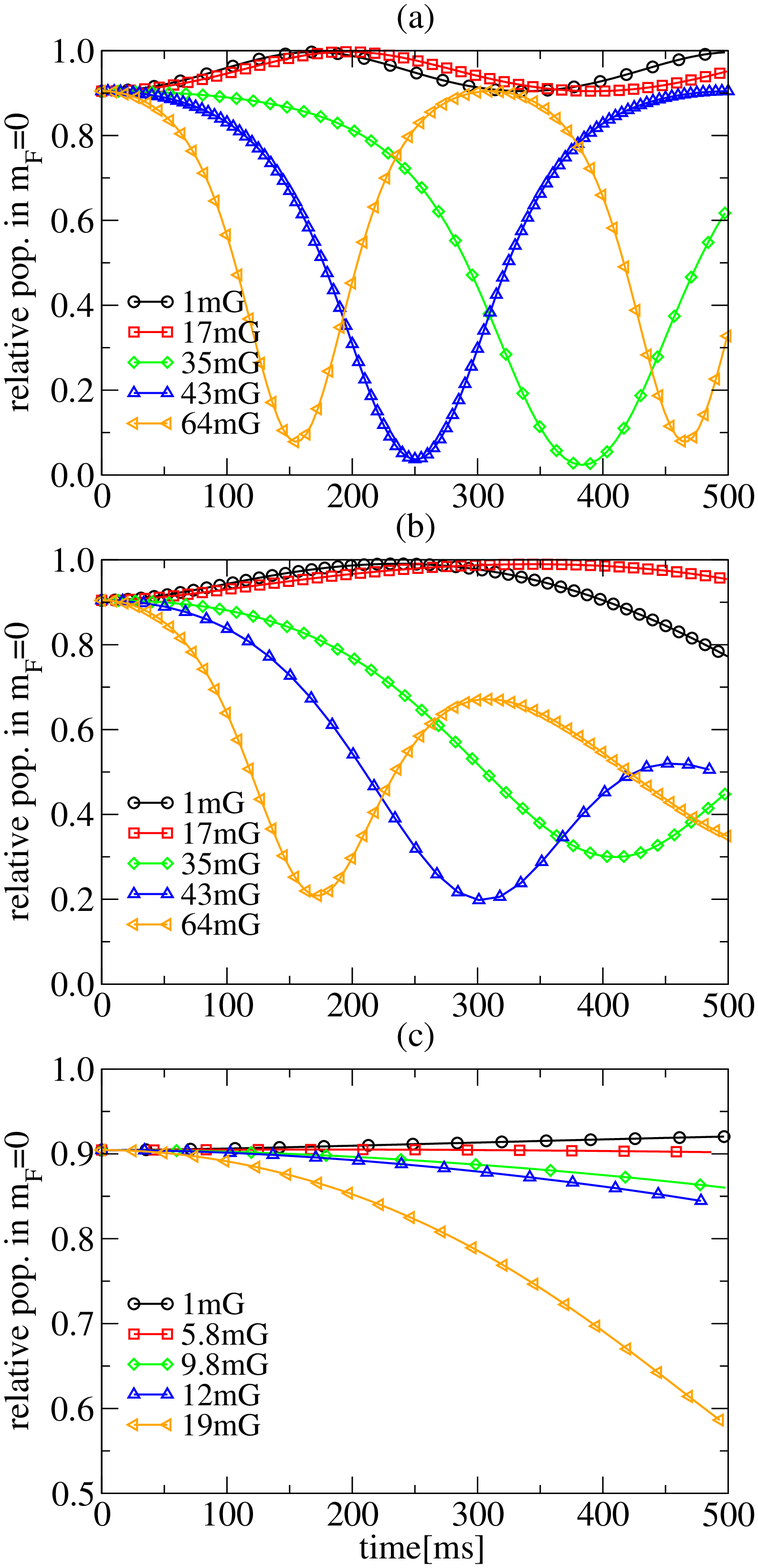}
\caption{(Color online)
Estimated time evolution of rubidium population at $|2,0>$ hyperfine spin
state when the quadratic Zeeman coupling or Zeeman
field varies.
(a) and (b) are for optical lattices with potential depth $V=10E_R$ and
(c) is for $V=6E_R$. In (a), we assume the density is uniform and there are no spin losses
while (b) and (c) are for traps with inhomogeneous density and a finite loss rate $1/\tau =(200ms)^{-1}$.
In (a),(b),(c),
magnetic fields
which yield
dynamical critical quadratic Zeeman coupling $q_{c1}( \sim q_{c2})$
are $35mG$, $35mG$
and $10mG$ respectively.
}
\label{fig-pop1}
\end{figure}

Dynamical stabilities of $|2,0>$ state below the threshold $q_{c1}$, or instabilities
of $(|2,2>+|2,-2>)/\sqrt{2}$ state below threshold $q_{c2}$
are directly induced by quantum fluctuations.
Coherent quantum dynamics around state $|2,0>$ ($(|2,2>+|2,-2>)/\sqrt{2}$)
below and above dynamical critical coupling $q_{c1}$($q_{c2}$)
are qualitatively different
and it is
therefore plausible to probe these dynamics that are driven mainly by quantum fluctuations
at relative low frequencies.
To explore this possibility, we have studied strongly damped population oscillations around a
condensate $|2,0>$ or a uniaxial state,
at different quadratic Zeeman couplings when the spin exchange loss time is set to
be $200$ms. For this part of calculations
we also take into account the density inhomogeneity of rubidium atoms in traps.
Main results are shown in Fig.\ref{fig-pop1} for different magnetic fields;
at a given field, the corresponding quadratic Zeeman coupling is given as $q_{B}=3(\mu_B B/2)^2/ 6.8 GHz$.

Our major findings are three-folded and outlined below.
Firstly, for a condensate with small deviations from $|2, 0>$ state,
because of fast spin exchange losses complete coherent oscillations around $|2,0>$ are hard to observe.
Secondly, dynamics can be dramatically modified by a finite quadratic Zeeman coupling.
When the coupling is much less than $q_{c1}$ (defined fo the density at the center of
a trap), dynamics become
slower when the quadratic Zeeman coupling
increases.  Beyond $q_{c1}$, dynamics become faster while the quadratic Zeeman coupling increases.
These behaviors are consistent with results for the case of uniform density shown in the previous
section; there, the oscillation  frequency becomes zero at a dynamical critical coupling.

\begin{figure}
\includegraphics[width=0.98\columnwidth]{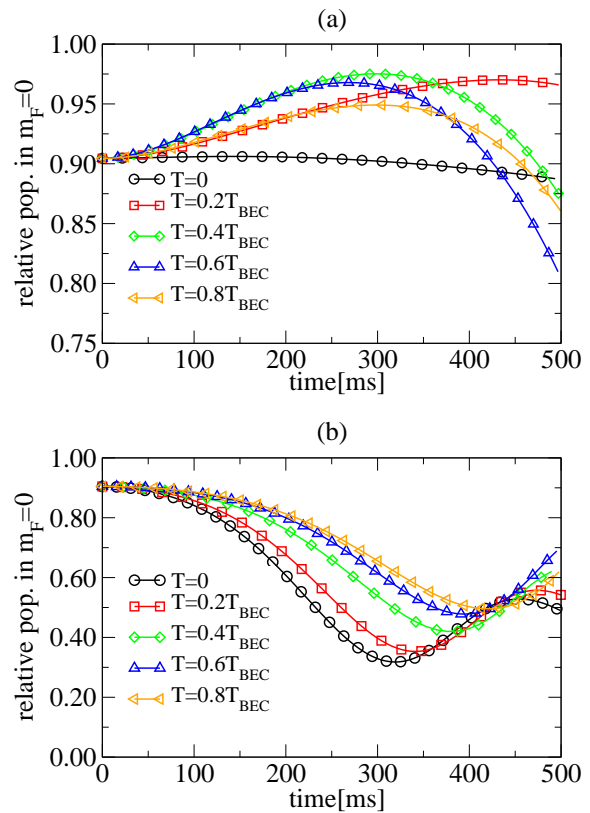}
\caption{(Color online)
Estimated time evolution of population at $|2,0>$ state at different
temperatures.
(a) is for a magnetic field of $13mG$ that is
below the dynamical critical field corresponding to $q_{c1}$ and (b)
is for $38mG$
field which is above $q_{c1}$.
The optical lattice potential depth is $V=8E_R$, $T_{BEC}=10.9nk$
and the spin exchange life time is again set to be $200ms$.
Density inhomogeneity in traps has also been taken into account to obtain
these plots.}
\label{fig-pop2}
\end{figure}

Thirdly (and perhaps most importantly), the population of atoms at state $|2,0>$ always {\em grows}
initially at short tome scales when the quadratic Zeeman coupling is zero or smaller than $q_{c1}$,
as a direct consequence of dynamical stabilities of a $|2,0>$-condensate. By contrast, beyond $q_{c1}$
the population of atoms at state $|2,0>$ always {\em decreases} initially at short time scales as a result of
dynamical instabilities in this limit.  These two distinct short-time behaviors of population at state
$|2,0>$ are signatures of  a transition from quantum-fluctuation driven dynamics to a mainly quadratic Zeeman coupling driven
dynamics. In the presence of strong spin losses, observing these distinct short-time asymptotics is an effective way
to probe many-body quantum fluctuations in condensates we have considered.
 At very large quadratic Zeeman coupling, we have found expected rapid oscillations
corresponding to mean field coherent dynamics and we don't show those results here.
Finally, we also estimate the time-dependence of population oscillations at finite temperatures
 using thermal-fluctuation induced potentials discussed in a previous section.
Thermal fluctuations usually speed up the coherent dynamics around state $|2,0>$ when the quadratic Zeeman coupling is
sufficiently weak (smaller than $q_{c1}$); on the other hand, they can also slow down coherent dynamics
around state $|2,2>+|2,-2>$ when a quadratic Zeeman coupling larger than $q_{c2}$ is present.
These results are shown in
Fig.\ref{fig-pop2}.
In obtaining these results, we have extrapolated the linear dynamical
analysis into a regime where oscillation amplitude is substantial.
So strictly speaking, results for $35mG$, $44 mG$, $65mG$ magnetic fields
shown in Fig.\ref{fig-pop1}(b)
are qualitative. However, short-time asymptotics that are the focuses of
our discussions here are accurate.

\section{Conclusion}

In conclusion, we have demonstrated that unlike the usual
mean-field-interaction driven coherent dynamics, {\em QFCSD} is a novel
class of coherent dynamics fully driven by quantum fluctuations.
These dynamics are conveniently tunable in optical
lattices where oscillation frequencies can be varied by three to four
order of
magnitude. Frequencies of oscillations driven by quantum
fluctuations
have both distinct lattice-potential dependence
and quadratic Zeeman-coupling dependence that can be studied experimentally.
For rubidium atoms, {\em QFCSD} can be directly
probed either at frequencies of a few Hz or when the
quadratic Zeeman coupling is about $10^{-1}nk$.
One of potential applications of QFCSD perhaps is the possibility of
performing precise measurements of strongly correlated quantum
fluctuations and
critical exponents near quantum critical points.
Current studies of critical correlations are based on analyzing
statistics of interference fringes\cite{Polkovnikov06,Hadzibabic06}.
Given the great control of coherent dynamics recently demonstrated for
cold atoms\cite{Obrecht07},
we would like to believe that quantum-fluctuation controlled dynamics
might be an alternative and promising path towards probing
critical correlations.
Finally, we have also investigated the effects of finite temperatures, spin exchange losses
and studied time evolution of condensates mainly driven by quantum fluctuations.

One of the authors(F.Z.) would like to thank KITP at Santa Barbara,
Institute of Physics, Chinese Academy of Sciences, CASTU at TsingHua
University in Beijing and Henry Poincare Institute in
Paris for their hospitalities in the spring and summer of 2007.
He also thanks Eugene Demler, Jason Ho,
Wolfgang Ketterle, Gordon Semenoff and Klaus Sengstock for valuable discussions on
experimental probing of {\em QFCSD}.
This work is supported by the office of the Dean of Science,
University of British Columbia, NSERC(Canada), Canadian Institute for Advanced
Research, and the Alfred P. Sloan foundation.

\appendix

\section{Collective coordinates}
To study QFCSD,
we expand $\psi^\dagger$ about a reference condensate $\chi$,

\begin{eqnarray}
\hat{\psi}^\dagger_{k, \alpha\beta} &=& \sqrt{4M}\chi_{\alpha\beta}
(\xi) \nonumber \\
&+& \frac{1}{\sqrt{N_T}}\sum_{\nu} L^\nu_{\alpha\beta} \big(
\hat{\theta}^\dagger_{\nu }(0)+
\sum_{{\bf q}\neq 0 } e^{i {\bf q}\cdot {\bf r}_k}
\hat{\theta}^\dagger_{\nu }({\bf q}) \big).
\label{Decomposition}
\end{eqnarray}
Here
$N_T$ is the number of lattice sites and $M$ is the number of atoms per site.
The superscripts (or subscripts) $\nu=x,y,z,\xi,p$ specify
condensate motion along the
five orthogonal directions of the submanifold.
$\hat{\theta}_\nu({\bf q})$ ($\hat{\theta}^\dagger_\nu({\bf q})$) are annihilation
(creation) operators of the collective excitations of mode $\nu$ with
lattice crystal momentum $\bf{q}$.
One can further include bilinear terms in the expansion (not shown
here).
The motion along these five orthogonal directions
is studied by introducing five mutually orthogonal matrices $L^\nu$, i.e.
$\textrm{Tr}(L^\mu
L^\nu)=2\delta_{\mu \nu}$.
Explicit forms of these orthogonal matrices
were introduced previously\cite{Song07}.

We are restricting ourselves to the dynamics of a condensate in a linear regime.
By expanding the original Hamiltonian in Eq.(\ref{Hamiltonian}) using the
decomposition introduced in Eq.(\ref{Decomposition}), we obtain an effective
Hamiltonian
that is bilinear in terms of $\theta^\dagger$ and $\theta$.
It is convenient to introduce five collective
Hermitian operators
$\tilde{X}_{\nu}$ defined in terms of $\theta^\dagger_\nu(0)$, $\theta_\nu(0)$ for ${\bf q}=0$ modes,
\begin{eqnarray}
\hat{X}_p&=&\frac{\theta^\dagger_p-\theta_p}{2i\sqrt{N}},
\hat{X}_t=\frac{\theta^\dagger_t+\theta_t}{2\sqrt{N}}, \nonumber \\
\hat{X}_x&=&\frac{\theta^\dagger_x+ \theta_x}{4\sqrt{N} (\chi_{yy} -\chi_{zz})},\nonumber \\
\hat{X}_y&=&\frac{\theta^\dagger_y+ \theta_y}{4\sqrt{N} (\chi_{zz} -\chi_{xx})},\nonumber \\
\hat{X}_z&=&\frac{\theta^\dagger_z+ \theta_z}{4\sqrt{N} (\chi_{xx} -\chi_{yy})}
\end{eqnarray}
where $N=N_T M$ is the total number of atoms in a lattice with $N_T$ sites.
By comparing Eq.(\ref{Decomposition})
and the expression for $\tilde{\chi}$ in terms of five spin coordinates
$X_\nu$ (see discussions before Eq.(\ref{xi-definition})), one
identifies
the semiclassical collective
coordinates $X_{\nu}$ as the expectation value of harmonic
oscillator operators $\hat{X}_\nu$.
In the following, we do not distinguish between $\hat{X}_\nu$ and $X_{\nu}$.
The conjugate momentum operators $\hat{P}_\nu$ can be
introduced accordingly so that the usual
commutating relations are obeyed, i.e
$[\hat{X}_\nu, \hat{P}_\mu]=i\delta_{\nu\mu}$.

{\bf Quantum-fluctuation-induced potentials}
Five modes of zero point motions (microscopic) around
$\chi(\xi)$ can also be labelled by the same set of indices.
For collective excitations of mode-$\nu$
with lattice momentum ${\bf q}(\neq 0)$,
creation operators are $\theta^\dagger_{\nu}({\bf q})$.
These operators obey the bosonic commutation
relations, $\left[ \theta^{~}_{k,\mu},
\theta^{\dagger}_{l,\nu}\right] = \delta_{kl}\delta_{\mu\nu}$.
Note that only the properties of three rotation modes and deformation mode ($\xi$-mode)
depend on the value of $\xi$.
The collective mode dispersion around $\xi=0$ or $\xi=\pi/2$ has a very
simple form,

\begin{eqnarray}
E_{\nu,\bf q}(\xi,X_\xi, q_B)=\sqrt{{\epsilon}_{\nu,\mathbf{q}}
[ 2m_{BN} v^2_{\nu}(\xi+X_\xi)+
{\epsilon}_{\nu, \mathbf{q}}]}.
\label{dispersion}
\end{eqnarray}
And in the expression for $E_{\nu,\bf q}$,
$\epsilon_{\nu, {\bf q}}=4 t_L \sum_\alpha (1-\cos q_\alpha d_L )+\kappa_\nu q_B$
is the energy of an atom with crystal quasi-momentum ${\bf
q}=(q_x, q_y, q_z)$;
$d_L$ is the lattice constant. $m_{BN}=1/(4 t_L
d_L^2)$ is the effective band mass.
$\kappa_\nu$, $\nu=x,y,z,\xi$
are coefficients introduced in
Eq.(\ref{effectiveH}),(\ref{mass}).
$v_\nu$ is the sound velocity of the $\nu$-mode in the small-$|{\bf q}|$
limit, $v^2_{\alpha}(\xi)={M}
\left(4b_LG_{\alpha\alpha}-c_L\right)/m_{BN}$,
$v^2_{\xi}={M} |c_L|/m_{BN}$.
$G_{\alpha\beta}(\xi)$ is a function of $\xi$ introduced in Eq.(\ref{mass}).

When the quadratic Zeeman coupling is zero, the energy dispersion was
also derived in a previous
work of ours\cite{Song07}.
The dispersion of the $\xi$-mode in this case is independent of $\xi$ or $X_\xi$.
And among four gapless spin modes, only the $x$-,
$y$- and $z$- spin rotational modes contribute to the $\xi$-dependence of the quantum fluctuation-induced potential
$V_{qf}$.
Substituting Eq.(\ref{dispersion}) into
Eq.(\ref{effectiveH}), one obtains the quantum-fluctuation-induced
potential
$V_{qf}$.
In 3d optical lattices, we find that
the $\xi$-dependent energy potential is
proportional to $\sum_\alpha v_\alpha^5/(d^5_L t^4_L) $,
$d_L$ is the lattice constant and $t_L$ is the hopping integral.
In 2d lattices,
the $\xi$-dependent energy potential $V_{qf}(\xi)$ is
proportional to $\sum_\alpha v_\alpha^4/(d^4_L t^3_L) \ln
t_Ld_L/v_\alpha$. These calculations lead to the scaling behaviors
discussed in Eq.(\ref{barrier}).
When the quadratic Zeeman coupling is present,
we find that for $\xi=0$,
$\kappa_{x,y,z,\xi}$ are not positive defined
and collective spin modes can be unstable.
To calculate induced-potential $V_{qf}$,
we include stable modes (${\bf q}\neq 0$) which
contribute to the renormalization of adiabatic condensate dynamics.
At $\xi=\pi/2$, all collective modes are stable.

{\bf Condensate depletion}
To study the condensate depletion in the optical lattice, we start from the
effective Hamiltonian. Five collective modes on top of a nematic state are
described by the following:
\begin{eqnarray}
{\cal H} = \sum_{\nu,q} E_{\nu,q}
\left(\tilde{\theta}^\dagger_{\nu,q}\tilde{\theta}^{}_{\nu,q}+\frac{1}{2}\right).
\end{eqnarray}
Here $\tilde{\theta}^\dagger_{\nu,q}$ and $\tilde{\theta}^{}_{\nu,q}$ are bosonic
creation and annihilation operators for the $\nu$-th mode. These operators can be
obtained by Bogoliubov transformations:
\begin{eqnarray}
\tilde{\theta}^{}_{\nu,q}
&=& u^{}_{\nu,q} \theta^{}_{\nu,q} + v^{}_{\nu,q} \theta^{\dagger}_{\nu,-q}, \\
\tilde{\theta}^{\dagger}_{\nu,q}
&=& u^*_{\nu,q} \theta^{\dagger}_{\nu,q} + v^*_{\nu,q} \theta^{}_{\nu,-q}.
\end{eqnarray}
The coefficients $u_{\nu,q}$ and $v_{\nu,q}$ are given:
\begin{eqnarray}
u_{\nu,q} &=& \frac{1}{\sqrt{2}}\left(\sqrt{1+\frac{m_{BN}v^2_{\nu, q}}{E_{\nu,q}}}+1\right)^{1/2} ,
\\
v_{\nu,q} &=& \frac{1}{\sqrt{2}}\left(\sqrt{1+\frac{m_{BN}v^2_{\nu, q}}{E_{\nu,q}}}-1\right)^{1/2}.
\end{eqnarray}
Condensate deletions from the $\nu$-th mode with crystal momentum $\bf q$ are
\begin{eqnarray}
\langle\theta^\dagger_{\nu,q} \theta^{}_{\nu,q} \rangle
= \left(u^2_{\nu,q}+v^2_{\nu,q}\right)
\langle\tilde{\theta}^\dagger_{\nu,q} \tilde{\theta}^{}_{\nu,q} \rangle
+ v_{\nu,q}^2 .
\end{eqnarray}
Here $\langle\tilde{\theta}^\dagger_{\nu,q} \tilde{\theta}^{}_{\nu,q} \rangle$ is
the occupation number of the $\nu$th mode and it is given by the Bose-Einstein
statistics:
\begin{eqnarray}
\langle\tilde{\theta}^\dagger_{\nu,q} \tilde{\theta}^{}_{\nu,q} \rangle
= \frac{1}{\exp\left(E_{\nu,q}/kT\right)-1} .
\end{eqnarray}
Putting all these together, we reach the final equation to determine the
condensate density:
\begin{eqnarray}
M - M_0 (T) =
\frac{1}{N_T}\sum_{\nu, q\neq0} \left(
v^2_{\nu,q} +
\frac{u^2_{\nu,q}}
{e^{\frac{E_{\nu,q}}{kT}}-1}
\right) .
\label{eqn-cf}
\end{eqnarray}
$M$ is the atom number density or number of atoms per lattice site and
$M_0$ is the condensed number density.
The first term in the summation gives quantum depletion number due to
the two body scattering and the second term gives the thermal depletion number.
We solve the above equation numerically and the
results are shown in Fig.\ref{parameters}.

At zero temperature, quantum depletion is given by the term
$v^2_{\nu,q}$ which
captures the effect of interactions or two body scattering processes.
Our estimate shows that
\begin{eqnarray}
1-\frac{M_0}{M}
&\sim&
\sum_{\nu=p,\xi,x,y,z}\frac{1}{M}\left(\frac{m_{BN}v_\nu}{t_Ld_L}\right)^3
\nonumber\\
&\sim& \frac{1}{M}\left(\frac{Ma_L}{t_L}\right)^{3/2}.
\end{eqnarray}
We have noticed that major contributions are from the
phase fluctuations since $a_L \gg |b_L|, |c_L|$.
In optical lattices when the potential depth $V$ increases, the bandwith $t_L$
decreases
exponentially as a function of $V$ and $a_L$ increases; as  a result, quantum
depletion grows as $(\frac{a_L}{t_L})^{3/2}$.

At finite temperatures, $M_0$ decreases as temperature $T$ increases and becomes zero at
the transition
temperature $T_{BEC}$. In the weakly interacting limit ($t_L>>a_L,b_L,|c_L|$), the
condensate
fraction is approximately equal to

\begin{eqnarray}
\frac{M_0(T)}{M_0(T=0)} = 1 - \left(\frac{T}{T_{BEC}}\right)^{\frac{3}{2}},
\end{eqnarray}
and $T_{BEC}\sim5M^{2/3}_0t_L$ in optical lattices.


\begin{thebibliography}{10}
\bibitem{stability}
State $|2,0>$ becomes unstable along the directions of $X_{x,y}$
in the presence of a finite $q_B$.
Coherent oscillations
discussed here can be observed
if initial states are prepared precisely.
More robust oscillations are those around states that correspond to
global potential minima
shown in Fig.\ref{fig-frequency0}b. See more discussions below.
\end{thebibliography}
\end{document}